\documentclass{article}
\usepackage{PRIMEarxiv}
\usepackage{framed}
\usepackage{siunitx}
\usepackage[english]{babel}
\usepackage[authoryear]{natbib}
\usepackage{amsmath}
\usepackage{mathtools}
\usepackage{comment}
\usepackage{graphicx}
\usepackage[colorlinks=true, allcolors=blue]{hyperref}
\usepackage{standalone}
\usepackage{wrapfig}
\usepackage{soul}
\usepackage{ulem}
\usepackage{authblk}
\usepackage{array, makecell}
\usepackage{multirow}
\usepackage{subscript}
\usepackage{xspace}
\usepackage{ifpdf}
\usepackage{ifxetex}
\usepackage{ifluatex}
\usepackage{mathtools}
\usepackage{bm}
\usepackage{graphicx}
\usepackage{tabularx}
\usepackage{textcase}
\usepackage{dashrule}
\usepackage{ragged2e}
\usepackage{authblk}
\usepackage{afterpage}

\usepackage{amsmath,amssymb,amsfonts}
\usepackage{algorithmic}
\usepackage{textcomp}
\usepackage{xcolor}
\usepackage{soul}
\def\BibTeX{{\rm B\kern-.05em{\sc i\kern-.025em b}\kern-.08em
    T\kern-.1667em\lower.7ex\hbox{E}\kern-.125emX}}

\usepackage{todonotes}
\usepackage{subfiles}
\usepackage{listings}
\usepackage{lscape}
\usepackage{makecell}
\usepackage{arydshln}
\usepackage{nicefrac}
\usepackage{tabularx,multirow,multicol}
\usepackage{natbib}
\setcitestyle{square}
\usepackage{graphicx}
\usepackage[linesnumbered,ruled,vlined]{algorithm2e}

\SetKwRepeat{Do}{do}{while}
\SetCommentSty{mycomm}

\newcommand{\figref}[1]{\figurename~\ref{#1}}
\newcommand{\tabref}[1]{Table~\ref{#1}}
\newcommand{\sectref}[1]{Section~\ref{#1}}
\newcommand{\algref}[1]{Algorithm~\ref{#1}}
\newcommand{\lineref}[1]{line~\ref{#1}}

\newcommand{\UAA}{A}
\newcommand{\manUAA}{\bar{A}}
\newcommand{\req}{r} 
\newcommand{\depl}{d} 
\newcommand{\rbt}{R}
\newcommand{\brPerRobot}{\lambda_\rbt} 
\newcommand{\nbPasses}{N_{p}}
\newcommand{\hPerWeek}{N_{h}}
\newcommand{\inBW}{W} 
\newcommand{\usedBWPart}{\tau} 

\newcommand{\maxBW}{W^{(M)}} 
\newcommand{\outBW}{\widehat{W}} 
\newcommand{\specEffUp}[1]{\upSpec_{#1}}

\newcommand{\upSpec}{\eta^{(up)}}

\title{To what extent can current French\\ mobile network support agricultural robots?\thanks{\textit{\underline{Citation}}: 
\textbf{La Rocca et al. 2024. 11th International Conference on ICT for Sustainability (ICT4S).}} }

\author[1]{Pierre La Rocca}
\author[1]{Gaël Guennebaud}
\author[1,2]{Aurélie Bugeau}
\affil[1]{Univ. Bordeaux, CNRS, Bordeaux INP, Inria, LaBRI, UMR 5800, F-33400 Talence, France}
\affil[2]{IUF}

\begin{document}
\maketitle

\begin{abstract}
The large-scale integration of robots in agriculture offers many promises for enhancing sustainability and increasing food production.
The numerous applications of agricultural robots rely on the transmission of data via mobile network, with the amount of data depending on the services offered by the robots and the level of on-board technology. Nevertheless, infrastructure required to deploy these robots, as well as the related energy and environmental consequences, appear overlooked in the digital agriculture literature. 
In this study, we propose a method for assessing the additional energy consumption and carbon footprint induced by a large-scale deployment of agricultural robots. Our method also estimates the share of agricultural area that can be managed by the deployed robots with respect to network infrastructure constraints.
We have applied this method to metropolitan France mobile network and agricultural parcels for five different robotic scenarios.
Our results show that increasing the robot's bitrate needs leads to significant additional impacts, which increase at a pace that is poorly captured by classical linear extrapolation methods.
When constraining the network to the existing sites, increased bitrate needs also comes with a rapidly decreasing manageable agricultural area.
\end{abstract}

\section{Introduction}
The integration of robots in agriculture offers many promises for improving sustainability and increasing food production. Agricultural robots are being developed for diverse applications such as optimized input use, sowing and harvesting automation and treating plant diseases \cite{OliveiraEtAl_AdvancesAgricultureRobotics_2021}. 
A large scale deployment of robots could reduce the drudgery of work, compensate for a lack of workforce and increase productivity \cite{OliveiraEtAl_AdvancesAgricultureRobotics_2021, MoysiadisEtAl_SmartFarmingEurope_2021, SparrowHoward_RobotsAgricultureProspects_2021, GerhardsEtAl_ComparisonSevenInnovative_2024}.
By taking advantage of innovation fields such as the Internet of Things or Artificial Intelligence, robots could further enhance the sustainability of the agricultural sector to limit the impact of agriculture on planetary boundaries while improving the resilience of agriculture against the effects of exceeding these limits \cite{CampbellEtAl_AgricultureProductionMajor_2017,RayEtAl_ClimateChangeHas_2019a}.

Several agricultural robot technologies rely on mobile networks. Real-Time Kinematic (RTK) correction improves robot positioning accuracy \cite{OliveiraEtAl_AdvancesAgricultureRobotics_2021}. RTK-correction based on mobile network overcomes the limited range of conventional radio systems and simplifies the management of frequency interference \cite{Orpheon_RobotsRTKGuidage_2024}. 

With the increasing deployment of autonomous robots, video streaming could ease their remote monitoring \cite{GreenEtAl_MeasurementLatencyRealtime_2021}. It could also open the door to their remote control \cite{RobertsPecka_4GNetworkPerformance_2018}.
Existing robots mainly process data on-board using low-power

Graphic Processing Units (GPU). However, there is a growing interest to externalize data treatments to datacenters~\cite{GSMA_SmartFarmingWeed_2020, RuigrokEtAl_ApplicationSpecificEvaluationWeedDetection_2020, AgroTIC_5GAgriculture_2021}. Among the expected benefits, companies could centralize data to provide better and personalized services. They could also centralize GPU, optimizing their use through scaling gains.
Removing embedded GPU could finally reduce robot cost while increasing their robustness and endurance as GPUs are sensitive and power hungry devices.

Existing studies attest to conclusive experiments of using the mobile network for agricultural robotics, both in terms of data rate \cite{RobertsPecka_4GNetworkPerformance_2018} and latency \cite{GreenEtAl_MeasurementLatencyRealtime_2021}. However, few concerns exist regarding a large-scale application of this process. 
In an experiment relying on a 5G standalone private network to convey multiple real-time video streams, the authors of \cite{ZhivkovEtAl_5GFarmEvaluating_2023} point out that this network configuration would not be sufficient to convey such a massive data flow over a long period of time. The authors of \cite{TomaszewskiEtAl_ApplicationMobileNetworks_2022} argue that the standards proposed for 5G in rural areas by the 3GPP\footnote{3GPP (3rd Generation Partnership Project): group of international organizations producing specification for cellular telecommunications technologies.} are theoretically not sufficient for most complex configurations. This includes robots and drones cooperation transmitting video flows to a distant server in real time. This limitation could open the door for more advanced networks such as an increased use of 5G private networks or even 6G in order to properly deploy most advanced robotic systems.

The energy consumption of mobile networks is well documented \cite{Coupechoux_Performances5GEtude_2021, GolardEtAl_EvaluationProjection4G_2023}. However it appears overlooked in the literature related to digital agriculture. This situation could reduce the potential benefits of agricultural robotics if their dependence on mobile networks remained ignored. For example, additional network uses could lead to an increase in network equipment, increasing energy consumption and greenhouse gas (GHG) emissions.
To fill this gap, we propose to study the potential consequences of a large-scale deployment of agricultural robots on mobile network usage and scaling. 
Our mobile network model is presented in \sectref{ntw}.

We model the deployment of robots in agricultural areas, taking into account the areas to be covered, robot capacities, network coverage and bandwidth constraints. Based on these specifications, we estimate the required network usage, compute the agricultural area manageable by the deployed robots and assess its consequences in terms of energy-related and manufacturing carbon footprints of the mobile network used by robots. Our methodology is detailed in \sectref{methodo}.

\begin{wrapfigure}{r}{0.5\textwidth}
    \includegraphics[width=0.9\linewidth]{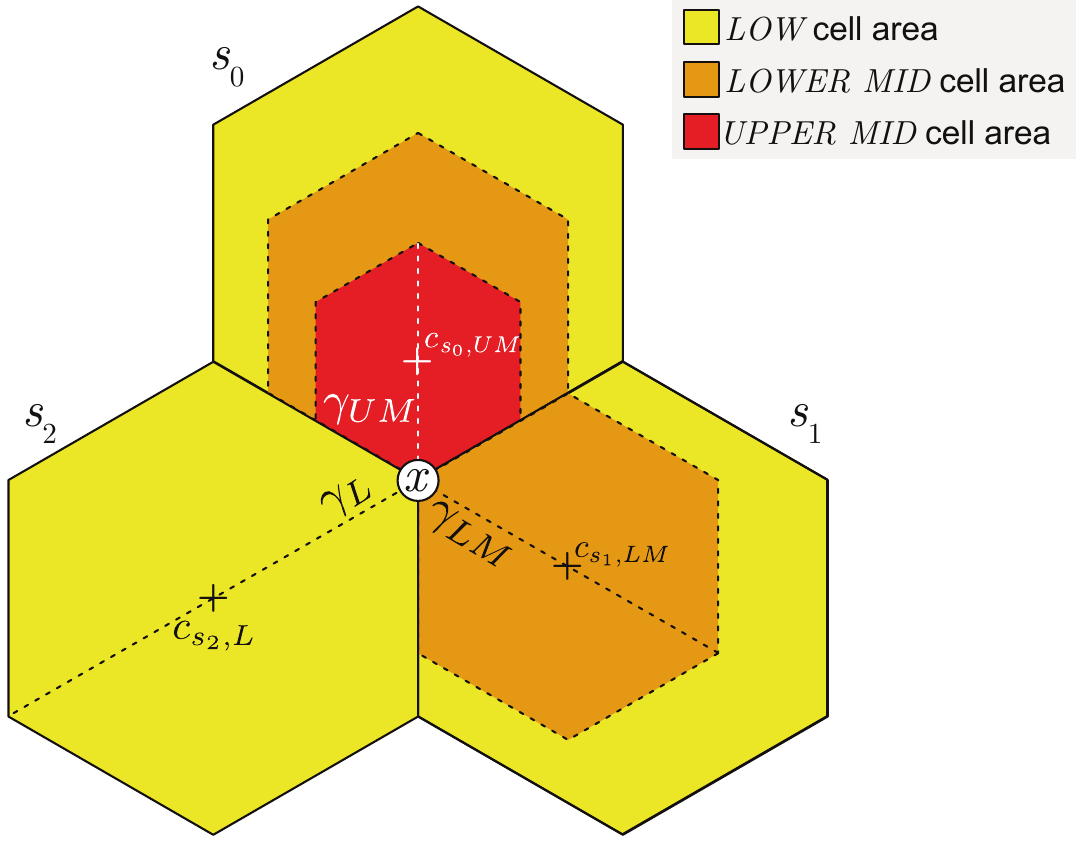}
    \caption{Graphical representation of a site $x$ with three sectors $s_1, s_2, s_3$ and their related cells. It illustrates possibly used cells for each sector among low, lower-mid (LM) and upper-mid (UM) frequency bands. The propagation radius $\gamma_{b}$ of a cell in the band $b$ corresponds to the diameter of the cell. The $\pmb{c}_{s,b}$ points denote the guessed centers of their respective cells (only three out of six are shown).}
    \label{fig:sect-scheme}
\end{wrapfigure}

A concrete implementation of our method is given in \sectref{results}.
We test two mobile network infrastructures in metropolitan France. The first is a fixed version of the existing French mobile network, restrained to a single operator.
The second is an upgraded version of the initial one where the existing sites are enhanced with additional frequencies to better satisfy the traffic demand from robots.

We test three main robot designs based on RTK-correction, remote monitoring through video streaming, and real-time edge computing. Each scenario corresponds to the deployment of homogeneous robots in a given territory. For the last twos, we distinguish two variants between a basic version and a high-definition one with higher bitrate needs, giving a total of five scenarios.
The results presented in \sectref{sec:disc} reveal highly heterogeneous impacts depending on the data traffic induced by the robots, both in terms of incremental energy consumption and the percentage of agricultural parcels that can be managed. 
The most demanding scenarios exhibit significant incremental energy consumption, while managing only a very small fraction of the agricultural parcels. These results question their relevance for large-scale applications.

\section{Mobile Network Model} 
\label{ntw}

This section presents our mobile network model, based on a few simplifying assumptions.
The effects of these main simplifications on our results will be discussed in Section~\ref{sec:disc}.

\paragraph{Sites}
    A mobile network consists of several individual sites, each associated with a georeferenced support (e.g., mast). In general, a site may be shared by several operators, each with its own base-station. A base-station refers to the operator's equipment (e.g., antennas and transmitters) enabling data transmission between end-user devices and the backhaul network. We here consider a single operator and use the term \textit{site} to refer to both the support and the unique base-station.

\paragraph{Sectors}
    As illustrated in \figref{fig:sect-scheme}, a site is typically composed of three non-overlapping sectors, radially oriented with an opening angle of 120°.
    Some sites may consist of just one or two sectors. The geographical extent of a sector corresponds to its largest cell, as defined below.

\paragraph{Cells}
    A sector $s$ consists of one or more overlapping cells. Each cell is characterized by a technology (4{G} or 5{G}), a frequency band $b$, and a spectral bandwidth $W_{s,b}$ (in MHz).
    Following~\cite{TheShiftProject_EnergieClimatReseaux_2024}, and for the sake of simplicity, we cluster frequency bands with similar properties into three aggregated bands. However, unlike~\cite{TheShiftProject_EnergieClimatReseaux_2024}, we further assume that each band is associated with a single technology, as follows:
    \begin{itemize}
        \item low (L): 4{G} 700 and 800 MHz bands;
        \item lower-mid (LM): 4{G} 1.8, 2.1 and 2.6 GHz bands;
        \item upper-mid (UM): 5{G} 3.5 GHz band.
    \end{itemize}

    Each band $b$ is associated with a propagation radius $\gamma_b$ (in km), which bounds the coverage area of the corresponding cell. As illustrated by \figref{fig:sect-scheme}, this radius decreases as the frequency band increases, causing the covered area of each cell of a sector to be nested in decreasing order.
    Since the low band provides the maximum coverage area, we impose that each sector includes at least one cell operating in the low band. Consequently, the extent of a sector is defined as the one of its low band cell.

\paragraph{Capacity model}
    A frequency band $b$ is associated with a spectral efficiency $\eta_b$ (in Mbps/MHz), which converts the available bandwidth $W_{s,b}$ of a given cell into its average download bitrate capacity (in Mbps).
    The available bandwidth $W_{s,b}$ bounded by a maximal bandwidth $\maxBW_b$, representing the sum of all sub bandwidths granted to the operator.
    To obtain the upload bitrate capacity, we apply a conversion factor $\rho_b$ reflecting the asymmetry between upload and download efficiencies: $\upSpec_b = \eta_b \rho_b$.
    The upload bitrate capacity of a cell is thus given by $\upSpec_b\,W_{s,b}$, while the capacities of overlapping cells within a sector can be aggregated to determine the total available capacity at a given location.

\begin{figure*}[t]
    \centering
    \includegraphics[width=\linewidth]{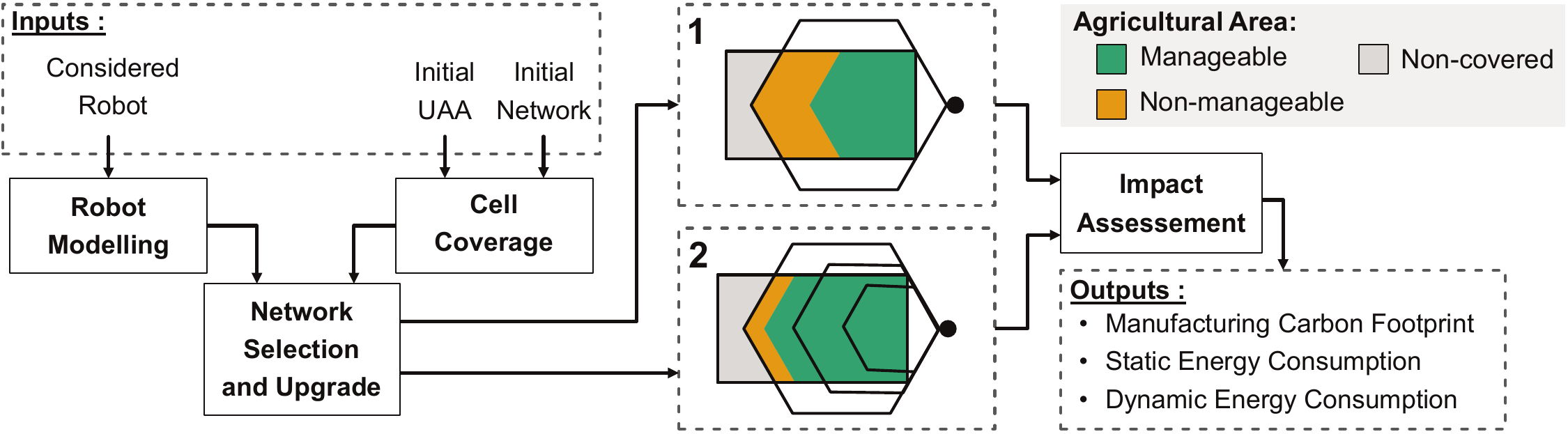}
    \caption{General overview of our assessment process. The \textit{Network selection and Upgrade} module simulates two robots deployments. \textbf{1}: based on the initial network;  \textbf{2}: based on an upgraded network version. Three types of agricultural areas are distinguished. Manageable: managed by robots, as the area is covered by mobile network with a sufficient bandwidth; Non-manageable: not managed by robots, as it is covered by mobile network but remaining bandwidth is not sufficient; Non-covered: not covered by mobile network.}
    \label{fig:gen-overview}
\end{figure*}

\paragraph{Network materiality and energy consumption}
    The network equipment inventory, along with the estimation of its carbon footprint and energy consumption, is borrowed from a previous study~\cite{TheShiftProject_EnergieClimatReseaux_2024}.
    Briefly, for each site, their model infers a list of equipment based on the list of cells that are present.
    While their model accounts for the yearly addition and renewal of equipment to maintain a heterogeneous infrastructure (in terms of power and manufacturing efficiency), ours assumes that all equipment has the same year of production, namely $2024$.
    The carbon footprint is then estimated using the \textit{stock} method, which calculates the sum of the embodied (manufacturing and transport) footprint of each piece of equipment normalized by its expected lifetime (in year).
    For the power consumption of an active equipment attached to a cell of band $b$, the authors of~\cite{TheShiftProject_EnergieClimatReseaux_2024} employs a simple power model consisting of two main parts:
    \begin{itemize}
        \item A \textit{static} power part $P_{static}^{(b)}$ corresponding to the energy consumption to keep the equipment switched on (24/7); 
        \item A \textit{dynamic} power part which is expected to be fully proportional to both the load (ratio between the traffic and maximal capacity, both in Mbps) and the bandwidth $W_{s,b}$ of the cell.
    \end{itemize}
    Since the maximum capacity is also proportional to $W_{s,b}$, the dynamic power part is more conveniently expressed as a function of traffic $\nu$ (in Mbps), leading to the following cell's power model:
    {
    \begin{equation}
        \label{eq:pcell}
        P_{cell}^{(b)} (\nu) = P_{static}^{(b)} + \frac{P_{dyn}^{(b)} \cdot \nu}{\eta_{b}^{(up)}}.
    \end{equation}}
    Here,  $P_{dyn}^{(b)}$ is given in W/MHz.
    At the site level, cell-based equipment is completed by active equipment that consumes a constant power $P_{site}$ per site, and $P_{b}$ per band present on the site.
    Integrating over a full year, the total yearly consumption $E_{tot}$ (in Wh) of the entire infrastructure is given by:
    {
    \begin{flalign}
    \centering
        \label{eq:etot}
        E_{tot} = \frac{8760}{(1-\sigma)} \sum_{x=1}^{N_{site}} \bigg( P_{site} + \sum_{b\in B_x} \Big(P_b + \sum_{s \in S_x} P_{cell}^{(b)}(\nu_{s,b}) \Big)\bigg)
    \end{flalign}
    }%
    where $N_{site}$ is the number of sites, $B_x$ (resp. $S_x$) the set of bands (resp. sectors) present on the site $x$, and $\nu_{s,b}$ the average traffic rate of the cell of band $b$ for sector $s$.
    Lastly, $\sigma$ accounts for overall losses, including power supply and AC/DC conversion.
    We refer to~\cite{TheShiftProject_EnergieClimatReseaux_2024} for details and values.

\section{Methodology}
\label{methodo}
We present our methodology for assessing the impact of a large-scale deployment of agricultural robots on the underlying mobile network infrastructure in terms of material requirements, carbon footprint, and energy consumption.

We refer to the input agricultural parcels as UAA for Utilized Agricultural Area.
We consider two network strategies:  
(1) an \textit{initial} version where the existing mobile network remains unchanged, and (2) an \textit{upgraded} version where additional bandwidth is deployed at existing sites to maximize the UAA that can be managed by robots.

For both versions, we simulate a deployment of robots, estimate the fraction of UAA that can be managed by those regarding network limitations, and assess the incremental carbon and energy footprint resulting from the robots' data traffic. For the upgraded version, we also account for the embodied carbon footprint of any new equipment.

\figref{fig:gen-overview} gives a general overview of our approach. 

The main inputs include a description of a generic robot model for seeding and weeding, the UAA as a raster map, and the initial network given as a list of sites and transmitters from a single operator.
A scenario is defined as these three entries, and assumes the deployment of a single robot model across the entire territory.

In this paper, we study the effect of the robot's workload and data rate properties as defined by the \textit{Robot modelling} module. Given a UAA sample (in hectares) and robots' properties, this module defines the number of required robots and their corresponding bitrate need.

\begin{figure*}[h!]
    \centering
    \includegraphics[width=\linewidth]{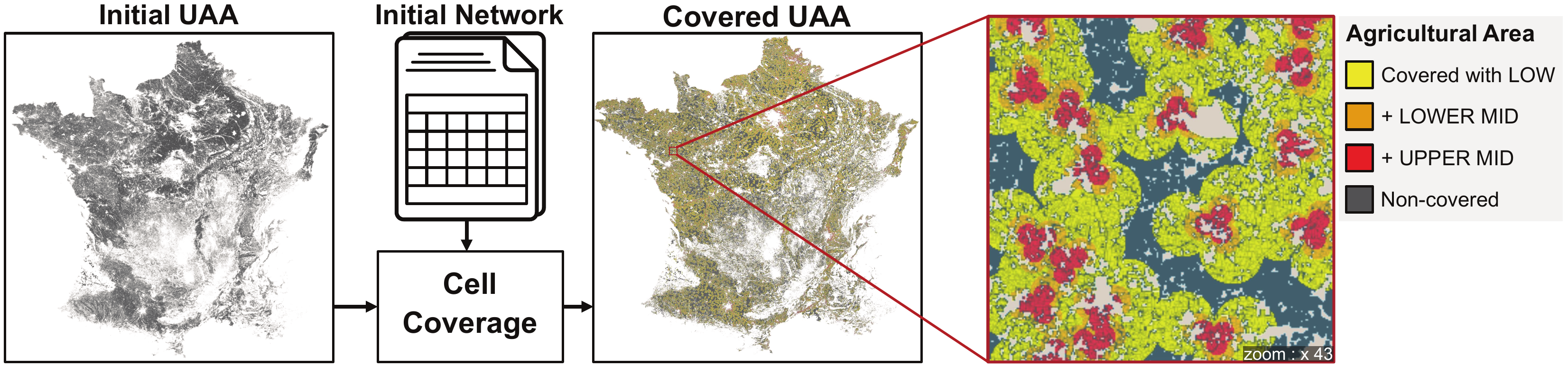}
    \caption{Zoomed agricultural area covered by frequency bands of the mobile network given by the UAA cover model. Selected UAA encompasses cereals, protein crops, oil-seeds, vines, forage and vegetables. Agricultural area displayed in yellow is covered by low, area displayed in orange is covered by low and lower-mid, area displayed in red is covered by low, lower-mid and upper-mid. Agricultural area in grey is not covered.}
    \label{fig:coverStep}
\end{figure*}

As depicted in \figref{fig:coverStep}, the \textit{Cell coverage} module processes the list of geo-referenced sites and transmitters to determine the extent of their respective sector and cell, which are then matched with the input UAA map to identify the portion of UAA covered by each cell.

The \textit{Network Selection and Upgrade} module is a key component of our methodology. For each sector, it estimates the number of required robots considering the covered UAA, and determines how many can actually be deployed, taking into account the possibly limiting bitrate capacity of cells.

A possible outcome of the first aforementioned network strategy is depicted as \textbf{1} in \figref{fig:gen-overview}. In this example, the number of robots required to manage the whole UAA within the sector implies a peak-rate demand exceeding the available capacity. As a result, only a fraction of the required robots can be deployed. Thus, only part of the underlying UAA is qualified as \textit{manageable} (in green in \figref{fig:gen-overview}).
Following the second \textit{upgrade} strategy, this module reduces the amount of \textit{non-manageable} UAA (in yellow) by either enlarging the spectral bandwidth of existing cells or by introducing additional cells with higher frequency bands.
A possible outcome of this strategy is depicted as \textbf{2} in \figref{fig:gen-overview},
where two cells have been added to increase the overall bitrate capacity, thereby increasing the extent of the manageable UAA.
Managing the remaining non-manageable and non-covered UAA (respectively shown in yellow and grey) would require the deployment of additional mobile network sites. However, this aspect falls beyond the scope of this study.

Lastly, the \textit{Impact Assessment model} performs a network equipment inventory of the sites and sectors serving deployed robots.
This step outputs the incremental embodied carbon footprint, as well as the static and dynamic energy consumption resulting from the network usage by the robots.
Following this overview, the four main modules of our approach are detailed in the next sub-sections.

\subsection{Robot modelling}
\label{rbtMdl}

A robot model $\rbt$ is defined by the following intrinsic properties:
a working width $l_\rbt$ (in metres), a working pace $v_\rbt$ (in m/s), and an upload bitrate need $\lambda_\rbt$ (in Mbps). We focus on the upload bitrate because it is the limiting factor in our studied scenarios.

We also define the extrinsic parameters $\nbPasses$ and $\hPerWeek$ representing the number of passes per year and the number of working hours per week, respectively. These extrinsic parameters are assumed to be common for all scenarios.
Assuming that the shortest period between two passes on the same parcel is one week, the maximum area that a single robot can manage is $u_\rbt = 0.36 \, l_\rbt v_\rbt \hPerWeek$ (in ha), where the factor $0.36$ converts $m^2/s$ to $ha/h$.
The number of robots required to manage an agricultural area $\UAA$ is $\left \lceil \frac{\UAA}{u_\rbt} \right \rceil$.
The fractional number of robots that can be simultaneously supported by a bitrate capacity $\Lambda$ is $\frac{\Lambda}{\brPerRobot}$.

\subsection{Cell Coverage}
\label{covModel}
This module determines existing network cells as well as the agricultural area they cover.
As illustrated by \figref{fig:sect-scheme}, each cell is associated with its \textit{guessed center} $\pmb{c}_{s,b}$ located at half its maximum propagation radius $\gamma_b$ along its sector's orientation (all cells within a sector share the same orientation). Using nearest-neighbour queries, each pixel of the UAA raster map is assigned to the cell of the closest $\pmb{c}_{s,b}$ within a distance $\gamma_b/2$.
To ensure consistency, pixels covered by multiple bands are assigned to cells within the same sector, prioritizing the highest frequency band. The UAA $\UAA_{s,b}$ covered by each cell is then computed by summing the values of its associated pixels.

This process is illustrated by the final image of \figref{fig:coverStep}, which displays the existing mobile network coverage for a zoomed-in agricultural area and for all frequency bands. It helps to understand the superposition of cells.

\subsection{Network Selection and Upgrade}
This step aims to identify which sectors' UAA (resp. robots) can be managed (resp. deployed) given the potential constraints of bitrate capacity.
It also selects the subset of initial cells needed to meet the robots' demands, and generates the so-called \textit{upgraded} network configuration by upgrading the capacity of existing sectors when necessary. These outputs serve as inputs for the next module, which assesses the incremental impacts.

The process is carried out independently for each sector $s$ by \algref{alg:deploymentAlg}.
In addition to previously introduced properties -- capacity $u_\rbt$ and upload bitrate $\brPerRobot$ for robots; covered UAA $\UAA_{s,b}$, existing and maximal bandwidths $\inBW_{s,b}$ and $\maxBW_{b}$ for cells -- this algorithm also takes as inputs $r_s = \left \lceil \nicefrac{\UAA_{s,L}}{u_r} \right \rceil$, the number of robots required to manage the UAA of $s$, and the percentage $\usedBWPart$ of the existing bandwidth assumed to be allocated to other usages at peak hours. For simplicity, this percentage is assumed constant for all sectors and cells. It also takes a Boolean value $upg$ to adapt its deployment behaviour and allowing upgrades of existing cells.
The algorithm outputs the number of deployable robots $d_s$, and the used cells recorded through their output bandwidths $\outBW_{s,b}$.

\newcommand{\capa}{c}
\newcommand{\rcapa}{p}
\begin{algorithm}
\caption{
Selection and Upgrade algorithm for a given sector $s$. 
}\label{alg:deploymentAlg}
\DontPrintSemicolon
\KwData{$\usedBWPart$, $u_\rbt$, $\brPerRobot$;\, $\forall b$, $\inBW_{s,b}, \UAA_{s,b} $, $\maxBW_{b}$; $\req_s$; $upg$}
\Begin{
    $\hat{\req} \gets \req_s$ \tcp*{\# remaining required robots}\label{line:reqrobot}
    $\depl_s \gets 0$ \tcp*{\# deployable robots}
    $\rcapa \gets 0$ \tcp*{\# remaining robots of prev. bands}
    $\forall b$, $\outBW_{s,b} \gets 0$ \tcp*{used bandwidths}
    $b \gets L$\;
    \While{$\hat{\req} > 0$ and $b \neq \emptyset$}{
        \label{line:addBand}
        $\capa \gets \nicefrac{(1-\usedBWPart) \inBW_{s,b} \specEffUp{b}}{\brPerRobot} + \rcapa$\;
        $\outBW_{s,b} \gets \inBW_{s,b}$\;
        \If{$upg$ and $\hat{\req}>\capa$}{
            $\outBW_{s,b} \gets \maxBW_{b}$\;
            $\capa \gets \nicefrac{(\maxBW_{b}-\usedBWPart \inBW_{s,b}) \specEffUp{b}}{\brPerRobot}+\rcapa$\label{line:upgrade}
        }
        $\Delta d \gets min(\left \lceil \hat{\req} \right \rceil ,\left \lfloor \capa \right \rfloor )$\label{line:upInit}\;
        $\depl_s \gets \depl_s + \Delta d$,
        $\rcapa \gets \capa - \Delta d$\;
        $n_{cres} \gets  \begin{cases} \hat{\req} - \nicefrac{\UAA_{s,b+1}}{u_\rbt}, & \text{if } b < UM \\
        0, & \text{otherwise} \end{cases}$\;
        $\hat{\req} \gets \hat{\req} - max(n_{cres}, \Delta d)$\;
        $b \gets \text{next}(b)$\;
    }
    \lIf{$\depl_s = 0$}{
            $\outBW_{s,b} \gets 0 ~|~ \forall  b$ \label{line:nonused}
    }
    \Return $\depl_s$, $\left\{\outBW_{s,b} ~|~ \forall  b\right\}$\;
    }
\end{algorithm}

After initializing output variables (lines 2~–~5), the algorithm simulates the deployment of robots by incrementally allocating the required robots' data traffic to cells, from low to upper-mid bands. 
The allocation stops either when all remaining required robots $\hat{\req}$ can be deployed using the visited bands, or when all bands have been considered (\lineref{line:addBand}).

For each cell of band $b$, the algorithm determines the available capacity $\capa$ in terms of the fractional number of robots that can be supported taking into account both the available bandwidth associated to $b$ and the remaining capacity $\rcapa$ coming from the previous bands. This assumes that a robot can simultaneously exploit the different available bands.

For the upgraded version, if the available capacity is insufficient to support $\hat{\req}$ robots, the bandwidth is upgraded to the maximum value $\maxBW_{b}$ and the capacity $\capa$ is updated accordingly (\lineref{line:upgrade}). If the existing cell is already at the maximal allowed bandwidth, this upgrade has no effect. Conversely, if $W_{s,b} = 0$, i.e. the cell does not exist, the operation reflects a cell addition.

The algorithm then computes the number $\Delta d$ of robots that can be deployed over the cell (\lineref{line:upInit}), and updates $\rcapa$. The number of remaining required robots is then updated using the maximum between $\Delta d$ and $n_{cres}$, the number of robots required for the area of the cell not covered by higher cells.
In the case where no robot can be deployed on the sector, we ensure the associated cells are properly reported as unused. (\lineref{line:nonused}).
From the algorithm's outputs, we can compute the amount of manageable UAA for each sector $s$, given by $\manUAA_{s,L} = \UAA_{s,L} \frac{\depl_s}{\req_s}$. 

\subsection{Impact Assessment} 
\label{envModel}

This last step primarily calls the network materiality and energy consumption model described in \sectref{ntw}.
In addition to the list of sites with their corresponding cells computed in previous steps, this model requires the per-cell average traffic rate $\nu_{s,b}$ to estimate dynamic power consumption induced by deployed robots as defined in \eqref{eq:etot}.
In our case, $\nu_{s,b}$ is computed for each scenario $R$ as:
$\nu_{s,b} = \frac{d_{s,b} \lambda_R k_R h_R}{8760}$,
where $d_{s,b}$ is the number of deployed robots per cell.
To compute the incremental energy and carbon impacts only, the general assessment method is adapted as follows. For the initial network configuration, the additional dynamic energy consumption is the one implied by the deployed robots.
For the upgraded configuration, in addition to the dynamic energy consumption, the static power consumption is computed as the difference between those of the upgraded and existing network configurations. For the embodied impacts, we explicitly enumerate the renewed RRUs (Remote Radio Units) as well as the BBUs (Broadband Base Unit), RRUs, and AAUs (Active Antenna Unit) of the added cells.

\section{Results}
\label{results}
We now present the application of the methodology to five robotic scenarios. All parameter choices and data used in the experiments are provided before we analyse the results.

\subsection{Scenarios, data and parameter choices}
\label{scenarios}

The definition of a scenario includes both territory-related and robot-related parameters, as presented below.
In this study, tested scenarios will only vary in terms of robot design, keeping all other parameters equals.
\newcommand{\mypar}[1]{\subsubsection*{#1}}
\subsection*{Territory}
Our methodology is applied to the French metropolitan territory, using the official carbon intensity factor of the 2023 electricity mix ($58$ gCO2e/kWh \cite{ADEME_Electricite2023Mix_2023}).

\mypar{UAA}
The data used for the UAA map comes from the French public agency for geographic and forestry information \cite{InstitutNationalDeLInformationGeographiqueEtForestiere_RegistreParcellaireGraphique_2022}. This data set provides details on the agricultural type of each parcel. 
We only retained parcels compatible with our type of robots, specifically those used for cereals, protein crops, oil-seeds, vines, forage and vegetables, representing a total of 14.6 million hectares. This represents 66\% of the total UAA given by the data set.
A rasterized version of the map was generated with a resolution of 232m per pixel, where pixel stores the underlying UAA in hectares.

\mypar{Initial Network}
\begin{wraptable}{r}{0.5\textwidth}
\renewcommand{\arraystretch}{1.2}
\caption{Default network parameter values \label{tab:hypNetwork}}
    \centering
\begin{tabular}{>{\centering\arraybackslash}m{26mm}*3{|>{\centering\arraybackslash}m{.07\textwidth}}}
\hline
                                                    & Low  & Lower-mid & Upper-mid \\ \hline
\textbf{Radius} ($\gamma_b$, km)                    & 4.5  & 2.25      & 1.5       \\ \hline
\textbf{Spectral Efficiency} ($\eta_b$, Mbps/MHz)   & 1.45 & 2.7       & 5.8       \\ \hline
\textbf{Upload/Download ratio ($\rho_b$)}           & \multicolumn{2}{c|}{0.25}      & 0.1\\ \hline
\textbf{Maximal Bandwidth} ($\maxBW_b$, MHz)        & 20   & 54.8      & 90        \\ \hline
\end{tabular}
\renewcommand{\arraystretch}{1}
\end{wraptable}
The data used for the initial network comes from the French public agency for radio-frequencies \cite{AgenceNationaledesFrequences_DonneesInstallationsRadioelectriques_2023}. We focused on a single operator, \textit{ORANGE}, which had the largest coverage at the time of data collection.
This open dataset provides site locations together with their list of cell types and bandwidths.

Our network model assumes that a sector contains  at least one cell with a low-frequency band and that the presence of a upper-mid (UM) cell implies the presence of a lower-mid (LM) cell.
While this holds for more than 96\% of sites, we observed that some urban sites had only upper-mid cells.
We thus consolidated this initial dataset by adding 3 000 low cells and 300 lower-mid cells (with the maximum bandwidths $\maxBW_b$), for a total of 80 000 low cells and 66 800 lower-mid cells.
The asymmetry between upload and download ratios was estimated from empirical measures \cite{Nperf_BarometreConnexionsInternet_2023}.
Following \cite{TheShiftProject_EnergieClimatReseaux_2024}, we assume that an average of $\tau = 20\%$ of existing bandwidth resources in rural areas are allocated to other usages. Default network parameter values are provided in \tabref{tab:hypNetwork}.
\subsection*{Robots' properties}
Robot designs were selected based on a literature review covering both academic research and industrial products \cite{Farmdroid_2023, PixelfarmingRobotics_2022, RuigrokEtAl_ApplicationSpecificEvaluationWeedDetection_2020}.
We chose use-cases reflecting both current and emerging technologies.
\begin{wraptable}{r}{0.5\textwidth}
\caption{Robots characteristics and implemented values}
    \centering
    \renewcommand{\arraystretch}{1.2}
    \begin{tabular}{>{\centering\arraybackslash}m{13mm}*5{|>{\centering\arraybackslash}m{.05\textwidth}}}
    \hline
     \textbf{Scenario} & \thead{\textit{RTK}} & \thead{\textit{Stream}} & \thead{\textit{HD}\\\textit{Stream}} & \thead{\textit{Edge}} & \thead{\textit{HD}\\\textit{Edge}}\\ \hline
     \textbf{Working Width} (m) & \multicolumn{3}{c|}{2}& \multicolumn{2}{c}{2} \\ \hline
     \textbf{Working Speed} (m/s) & \multicolumn{3}{c|}{1.25} & \multicolumn{2}{c}{1}\\ \hline
     \textbf{Workload} (ha/h) & \multicolumn{3}{c|}{0.9} & \multicolumn{2}{c}{0.72}\\ \hline
     \textbf{Required bitrate} (Mbps) & $0.012$ & 1 & 3  & 9 & 25 \\ \hline
    \end{tabular}
    \renewcommand{\arraystretch}{1}
    \label{tab:robots}
\end{wraptable}

For all scenarios, robots are assumed working 8h per day, 5 days a week, with, in the worst-case, a single week between two passes on the same parcel. 

This assumption only accounts for active fieldwork, excluding battery charging and travel times between charging stations and parcels, or between parcels.
Robots are assumed to perform $\nbPasses=12$ passes per year: two sowing passes plus an average of ten for weed treatments and other mechanical labour as reported in \cite{Agreste_EnquetePratiquesCulturales_2024}.
This average number corresponds to a scenario where mechanical weeding fully replaces chemical-based weeding, this substitution being one of the main arguments in favour of advanced agricultural robots \cite{GerhardsEtAl_ComparisonSevenInnovative_2024, OliveiraEtAl_AdvancesAgricultureRobotics_2021, RuigrokEtAl_ApplicationSpecificEvaluationWeedDetection_2020}.
Robot variants are summarized in \tabref{tab:robots} and presented below.

\mypar{RTK}
Real-Time Kinematic correction improves positioning accuracy and is widely used in agricultural machinery.
It requires the robot to continuously send its geographical position to a server, which then computes and returns a corrected position. According to \cite{Orpheon_CentreCalculOrpheon_2018}, between 1 and 3 positions are exchanged per seconds between the robot and the server.
Based on \cite{Farmdroid_2023}, the average download bitrate for RTK is around $12$ kbps. This is approximately 10 times higher than the upload bitrate. Unlike other scenarios, we conservatively retain this download average bitrate for the \textit{RTK} scenario. Given its low traffic rate, this scenario mostly serves to assess the coverage ratio rather than network capacity constraints.

\mypar{Stream}
Video streaming enables remote supervision of robots and is frequently integrated into robotic solutions by manufacturers~\cite{Farmdroid_2023, PixelfarmingRobotics_2022}.

This scenario assumes all robots continuously transmit a video feed to servers or farmers for monitoring.
The default variant assumes a medium-quality stream at 1 Mbps (720p at 20 fps with the H264 codec), while the \textit{HD Stream} variant relies on a 1080p video stream at 3.8 Mbps~\cite{_ZoomSystemRequirements_}. Higher video quality could improve the remote control.

\mypar{Edge}
Edge-based decision involves moving embedded computing units from robots to edge servers \cite{Marwala2024}. 
This design is quite prospective and is more a matter of academic or industrial research communication
\cite{RuigrokEtAl_ApplicationSpecificEvaluationWeedDetection_2020, GSMA_SmartFarmingWeed_2020} than of a commercialized product. 
Applying edge computing to agricultural robots could reduce their onboard processing requirements, instead relying on datacenters with pooled computing resources.
It requires the robot to send high-quality video streams to a edge-server, analysing the videos to detect plants, and sending back its response for the robot to act. As this should be done in real time, latency must be lower than 250 ms, justifying the use of edge servers rather than centralized cloud datacenters.
The required bitrate is estimated at 9 Mbps based on \cite{RuigrokEtAl_ApplicationSpecificEvaluationWeedDetection_2020}, assuming a 2m working width with 3 RGB cameras capturing crop rows at 4 fps.
This case requires a slightly reduced robot speed to allow sufficient time for data transmission, processing, and response.
The \textit{HD Edge} variant assumes a higher quality video stream at 25 Mbps, aligning more closely with the empirical findings by Ruigrok et al. \cite{RuigrokEtAl_ApplicationSpecificEvaluationWeedDetection_2020}, which strove to achieve peak rates of 120 Mbps.

\begin{table}[]
    \caption{Total number of cells in the initial network and those used by deployed robots.
    }
    \label{tab:existingSectors}
    \centering
\begin{tabular}{l|c|c|c} \hline
Number of cells     & Low       & Lower-mid     & Upper-mid \\ \hline
Full Network        & 84 631    & 66 832        & 27 213 \\ 
Covering UAA        & 61 381    & 43 744        & 13 418  \\ \hline \hline
Used by RTK         & 61 381    & 0             & 0     \\ 
Used by Stream      & 61 381    & 19 862        & 0     \\ 
Used by HD Stream   & 58 310    & 34 520        & 148   \\ 
Used by Edge        & 43 781    & 43 744        & 3 428 \\ 
Used by HD Edge     & 19 868    & 19 868        & 8 052 \\ \hline
\end{tabular}
\end{table}

\begin{table}[]
    \caption{Cell count breakdown for the upgraded network.
    }
    \label{tab:upgUpdSectors}
    \centering
    \newcommand{\initcells}{Initial cells}
    \newcommand{\upcells}{Updated cells}
    \newcommand{\addedcells}{Added cells}
    \newcommand{\cead}[1]{\multicolumn{1}{c|}{#1}}
    \begin{tabular}{l|l*3{|r}} \hline
    Scenario                        &           & \cead{Low} & \cead{Lower-mid} & \cead{Upper-mid} \\ \hline
    \multirow{3}{*}{RTK}            & \initcells   & 61 381     & 0             & 0 \\ 
                                    & \upcells     & 0          & 0             & 0 \\ 
                                    & \addedcells  & 0          & 0             & 0 \\ \hline \hline  
                                    
    \multirow{3}{*}{Stream}         & \initcells   & 42 813     & 19 859 & 0 \\ 
                                    & \upcells     & 18 568     & 3             & 0  \\ 
                                    & \addedcells  & 0          & 7 105         & 0 \\ \hline 
                                    
    \multirow{3}{*}{HD Stream}      & \initcells   & 34 315     & 27 890        & 148 \\ 
                                    & \upcells     & 27 066     & 6 630         & 0  \\ 
                                    & \addedcells  & 0      & 11 483        & 0  \\ \hline  \hline 
                    
    \multirow{3}{*}{Edge}           & \initcells   & 30 917     & 23 907        & 3 428 \\
                                    & \upcells     & 28 189     & 19 837        & 0 \\ 
                                    & \addedcells  & 0     & 15 354        & 18 006 \\ \hline
                            
    \multirow{3}{*}{HD Edge}        & \initcells   & 30 913     & 17 184        & 8 052  \\
                                    & \upcells     & 28 189      & 26 560         & 0  \\ 
                                    & \addedcells  & 0     & 15 358        & 33 988   \\ \hline
    \end{tabular}
\end{table}

\begin{table}[t]
    \caption{Environmental results obtained for full network, network covering UAA, and network used by each scenario. 
    The results for the scenario are the delta compared to the full and covering UAA networks.
    The column intensity based estimations present results from a naive linear extrapolation. EC : Energy Consumption. CF : Carbon Footprint.}
    \label{tab:envRes}
    \begin{tabular}{l|c|rr|cc|cc|cc||cc} \hline
    \thead{Scenario}   
    &\thead{Network} 
    &\multicolumn{2}{c|}{\thead{Manageable\\UAA\\+ depl. robots (\%)}}  
    &\multicolumn{2}{c|}{\thead{Total EC \\(GWh)\\ $[$Dyn. Part (\%)$]$}}  
    &\multicolumn{2}{c|}{\thead{Manuf.\\ CF\\(ktCO2e/y)\\$[$Renewal Part (\%)$]$}}  
    &\multicolumn{2}{c||}{\thead{Total CF \\(ktCO2e/y)\\$[$Manuf. Part (\%)$]$}} 
    &\multicolumn{2}{c}{\thead{Intensity\\based\\estimations \\(GWh,tCO2e)}} \\\hline\hline
    
                      & Full Network                   &   &                & 653.7 & $[38\%]$  & 47    &       & 84.9  & $[$55\%$]$ & 754 & 84 \\ \cdashline{2-12}
                      & Covering UAA                   &  &                & 424.3 & $[35\%]$  & 33.7  &       & 58.3  & $[$58\%$]$ & &  \\ \hline 
    
    \multicolumn{12}{c}{\multirow{2}{*}{\large$+$}}\\
    \multicolumn{12}{c}{}\\\hline
    RTK                         & Existing      &  \hspace{6mm} 83\% & $100\%$ & $< 0.01$   &   & 0  &   & 0  &  &  0.2  &  0.02 \\ \cline{1-12}
        
    \multirow{2}{*}{Stream}     & Existing      &  53\%  & $64\%$ & 5.4 &             & 0 &           & 0.3  &            & 11.1 & 1.2 \\ 
                                & Upgraded      &  70\%  &$84\%$ & 64.1 & $[14\%]$    & 3.6 & $[56\%]$ & 7.3  & $[49\%]$   & 14.6 & 1.6 \\ \cline{1-12}
        
    \multirow{2}{*}{\thead{HD\\Stream}}  & Existing      &  32\%  &$39\%$ & 11.1     &             & 0 &           & 0.6  &            & 18.8 & 2.1 \\ 
                                & Upgraded      &  52\%  &$63\%$ & 123.7   & $[17\%]$    & 7.4 & $[55\%]$ & 14.7  & $[50\%]$   & 30.8 & 3.4 \\ \cline{1-12}
    
    \multirow{2}{*}{Edge}       & Existing      & 13\%  &$15\%$   & 21.8&           & 0 &           & 1.3  &            & 25.4 & 2.8 \\ 
                                & Upgraded      & 38\%  &$46\%$   & 290 & $[22\%]$  & 18.9 & $[34\%]$ & 36.4  & $[52\%]$  & 75.4 & 8.4  \\ \cline{1-12}
        
    \multirow{2}{*}{\thead{HD\\Edge}}    & Existing      & 4\%   &$5\%$   & 14.8   &             & 0 &           & 0.9  &            & 24.5 & 2.7 \\ 
                                & Upgraded      & 23\%  &$28\%$  & 454.4  & $[15\%]$    & 36.2 &$[20\%]$& 64.4 & $[56\%]$   & 128.9 & 14.4 \\
    \hline                  
    \end{tabular}
\end{table}

\subsection{Analysis}
\label{Analysis}

We now present and analyse the results obtained for crop-related agricultural areas of the French territory.

\tabref{tab:existingSectors} reports cell statistics per frequency band and scenarios when using the initial network only.
The first two rows indicate the total number of cells in the initial network and, among them, the ones covering our targeted parcels.
Following rows indicate for each robotic scenario the number of cells used by deployed robots when allocating the cells from lowest to highest frequencies by \algref{alg:deploymentAlg}. 
Remark that in this configuration the more data-intensive a scenario, the fewer the number of low cells used. This is because some existing sectors only have a low cell with limited bandwidth, thus cannot even support a single robot at such high bitrates.
Remark that none of our scenarios make use of the 13 418 existing upper-mid cells covering agricultural areas. At most, 60\% of them is being exploited. This is because most of them are rather located in urban or peri-urban area, hence covering a very small fraction of UAA and making them unusable for our purposes.

\tabref{tab:upgUpdSectors} reports the number of cells per frequency band for the upgraded network configuration. Cell quantities are decomposed into three parts: unchanged cells of the existing network, updated cells (enlarged bandwidth), and added new cells. For instance, considering the \textit{Edge} scenario, the initial 43 744 lower-mid cells covering UAA (\tabref{tab:existingSectors}) are decomposed into 23 907 unchanged and 19 837 updated cells in this table.

As expected, the \textit{RTK} scenario only relies on low frequency cells as its data requirements are minimal.
In contrast, the \textit{Stream} and \textit{Edge} scenarios make a much heavier use of lower-mid cells and, with the \textit{Edge} scenarios even requiring upper-mid cells to meet their high data needs.
The HD and non-HD variants exhibit significant differences too.
These are clearly visible in both the initial cell uses (\tabref{tab:existingSectors}), or the cell updates and additions in the upgraded network (\tabref{tab:upgUpdSectors}).
For instance, the \textit{HD Stream} and \textit{Edge} scenarios imply nearly that all involved sectors be equipped with both low and lower-mid cells at full bandwidth, either by updating initial cells that are not already complete, or by adding new cells.
In the \textit{Edge} scenarios the numbers of newly added upper-mid cells are substantial, reaching approximately 66\% to 125\% of the total existing upper-mid cells across the entire territory.
To better understand the meaning of these numbers it is crucial to also take into account the respective percentage of the manageable UAA reported in the third column of \tabref{tab:envRes}.
According to our model, about 83\% of the UAA is covered with 4G-low.
However, the percentage of managed UAA drastically varies across the \textit{Stream} and \textit{HD Edge} scenario decreasing from 53\% to 4\% only when using the initial network, and from 70\% to 23\% after upgrading existing sites.

Overall, increasing the UAA that can be managed requires a significant amount of equipment renewal or addition.
Even the least traffic-intensive \textit{Stream} scenario necessitates 7 108 lower mid cell renewals or additions. Unlike the \textit{Stream} scenarios which do not heavily depend on upper-mid cells, the \textit{Edge} scenarios appear hardly feasible without a massive deployment of 5G in the 3.5GHz band all over the rural areas.

In addition to UAA and robots percentages, \tabref{tab:envRes} also reports energy consumption and carbon footprint results. The first two rows respectively provide reference values for the full network and the sub-part covering agricultural area.
Their dynamic energy consumption has been computed for an average load of 20\%, a figure consistent with the prior assumption of allocating 20\% of the existing bandwidth to other usages.

The energy consumption estimate for the full network appears to be about $80\%$ lower than the one reported by Orange~\cite{RouphaelEtAl_ImpactNetworksGreenhouse_2023}. Reasons are twofold. First, we consider 4G and 5G cells only, ignoring GSM and UMTS cells. Second, we assume up-to-date equipment that benefits from the latest advances in energy efficiency, whereas the actual network comprises a mix of older and newer equipment. Overall, this difference does not affect the way our model deploys robots nor the general conclusions we could make from our results.

The scenario rows report only the incremental energy consumption and carbon footprints, as explained in \sectref{envModel} for the two network configurations.
For instance, since the \textit{RTK} scenario does not require any updates to existing sites, its incremental manufacturing footprint vanishes. Additionally, given its relatively low data traffic, its incremental dynamic power consumption is negligible too.
The carbon and energy results align with the previously discussed cell counts. We observe a significant variation in magnitude orders depending on the scenario.
The higher the bitrate required by the robot, the lower the managed UAA and the greater the incremental carbon footprint.
Notably, the carbon footprint associated with energy consumption increases in more data-intensive scenarios, consistent with the rise in dynamic energy consumption. 

Under our hypotheses, the incremental carbon footprint is evenly distributed between electricity consumption and embodied emissions.
Comparing these results to the reference values of the full or UAA-related portion of the existing network highlights their high scale. For instance, the two \textit{Edge} scenarios would lead to a $44\%$ and $70\%$ increase of the total energy consumption of the existing network while leaving about $60\%$ and $80\%$ of the non-manageable UAA, respectively.

In terms of embodied emission, the increase is less pronounced. This is because we upgrade existing sites only, thus benefiting from the existing equipment that can be shared with the renewed/added cells (e.g., support, shelter, heptaband antennas, some BBU components, etc.).

\subsection{Comparison to intensity based extrapolation}
To demonstrate the relevance of our approach, we compare our results to a naive linear extrapolation solely based on overall traffic volume (in GB) multiplied by factors expressed per GB of transferred data. For energy consumption, we used a factor of 224 Wh/GB obtained by the ratio between the 2022 electricity consumption of the mobile network in metropolitan France~\cite{ARCEP_enquete_annuelle_2024} and the corresponding mobile traffic~\cite{ARCEP_observatoire_2024}.
Both values correspond to real data reported by the operators.
For the carbon footprint, we directly used the 25 gCO2e/GB factor from the \textit{Base IMPACTS}~\cite{ADEME_negaoctet_2023} which is representative of the same territory. These extrapolations are reported in the last two columns of \tabref{tab:envRes}.

For the \textit{Full Network} reference, we considered 29\% of the total annual traffic~\cite{ARCEP_observatoire_2024} to match the market share of the operator we selected.
This comparison for the \textit{Full Network} is in-line with the already noticed and expected underestimation of its energy consumption by our model, but both indicators (energy and carbon footprint) are still close enough to assess the faithfulness of the bottom-up inventory and environmental assessment model we used.

However, this naive extrapolation completely fails to provide a faithful estimation of the true incremental impacts induced by our scenarios, both in terms of energy consumption and carbon footprint.
Such naive extrapolations are not only off in terms of orders of magnitude, but they also fail to correctly capture the relative variations. These results validate the importance of considering the equipment inventory induced by the addition of new uses in order to better anticipate the overall consequences on the environment.

\subsection{Sensitivity Analysis}
\begin{wrapfigure}{r}{0.4\textwidth}
    \centering
    \includegraphics[width=\linewidth]{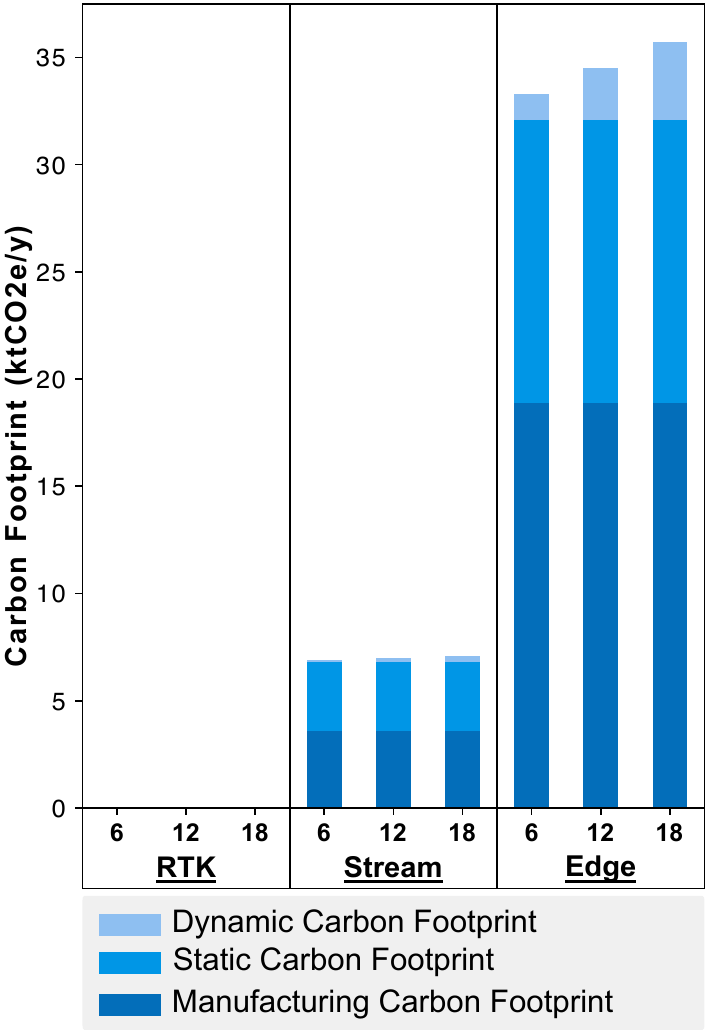}
    \caption{Additional carbon footprint for RTK, Stream and Edge scenarios when varying the number of passes annually done by robots, between 6, 12 and 18.
    }
    \label{fig:sensitivityAnalysisWW}
\end{wrapfigure}
\subsubsection{Number of passes}

As our model relies on several parameters, small variations in these parameters may affect overall results.
Since the previous results have already explored the variability along the robot's bitrate need, this section focuses on the sensitivity of other parameters.
For clarity, we solely focus on the upgraded network configuration and leave out the HD variants.

\figref{fig:sensitivityAnalysisWW} shows the effect of a $\pm 50\%$ variation of the number of passes per year $\nbPasses$ on the yearly carbon footprint for each scenario. This parameter has no effect on the number of upgraded cells, and it thus exhibits a rather small overall influence with at most a 10\% difference between the two extreme values for the \textit{Edge} scenario. In terms of carbon footprint related to energy consumption, this difference raises to $15\%$.

\subsubsection{Network parameters}
\begin{figure*}[t]
    \centering
    \includegraphics[width=\linewidth]{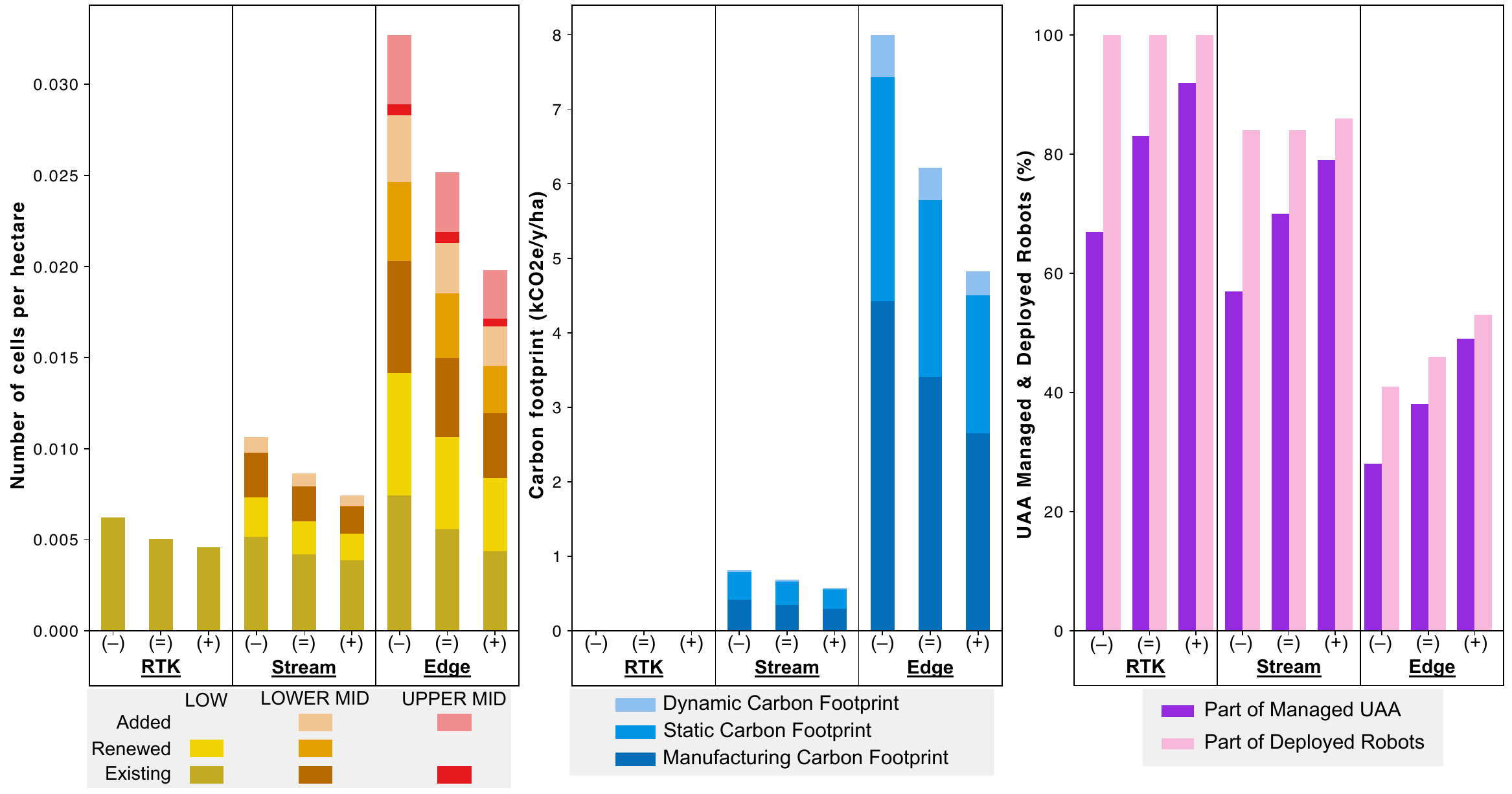}
    \caption{Number of cells per hectare, additional carbon footprint per hectare and part of UAA managed and deployed robots for RTK, Stream and Edge scenarios when jointly varying by 20\% the spectral efficiency and propagation radius of frequency bands. The pessimistic variation is denoted as ($-$), optimistic variation as ($+$) and the initial set of parameters as ($=$).
    }
    \label{fig:sensitivityAnalysisNetwkSites}
\end{figure*}

\figref{fig:sensitivityAnalysisNetwkSites} illustrates the effects of a $\pm 20\%$ variation in both the spectral efficiency $\eta_b$ and the propagation radii $\gamma_b$ on three indicators: cell count per hectare, incremental carbon footprint per hectare and the percentage of manageable UAA.
Here, the use of per-hectare indicators better reflect the relative effect of each parameter.
Increasing the propagation radius enables a cell to cover a larger UAA, while increasing the spectral efficiency implies that a cell can convey more data for a same bandwidth. The optimistic (resp. pessimistic) results correspond to an increase (resp. decrease) of both parameters.
The direct effect of these parameters is, for a limited quantity of cells and equipment, to substantially reduce or increase the amount of manageable UAA.
Normalizing the absolute incremental impacts by the manageable UAA, thereby indirectly leads to substantial variations of the per hectare cell count and carbon footprint.
The combined effect of the $\pm 20\%$ variation of these two input parameters remains contained as we observe a $\pm20\%$ and $\pm30\%$ variation of the per hectare carbon footprint for the \textit{Stream} and \textit{Edge} scenarios respectively.
We also note that once normalized, the relative footprint difference between the \textit{Stream} and \textit{Edge} scenarios is much more important ($\times9.1$) than what it seemed to be when only looking at the absolute values ($\times5$, \tabref{tab:envRes}).

\section{Discussion}
\label{sec:disc}
In this section, we put our results in regards to the impacts of the robot themselves, before discussing some design choices intrinsic to our models and more general aspects.

\subsection{Robot impacts}
This study focused on assessing the energy and carbon footprint of a sub-part of the underlying mobile network, while excluding the footprint of other involved equipment and services such as installation and maintenance, the servers, and the robots themselves.
For the latter, the authors of~\cite{LaRoccaEtAl_EstimatingCarbonFootprint_2024} estimated the carbon footprints of large-scale deployments of similar devices. By adjusting their findings to align with our assumptions on the number of passes and workload, we estimate a rough order of magnitude of about 280 and 140 GWh/year for the deployed robot power consumption of the \textit{Stream} and \textit{Edge} scenarios respectively. Comparing these estimates to the respective energy consumption of the upgraded network (64 and 290 GWh/year from \tabref{tab:envRes}), it becomes evident that the network energy consumption cannot be ignored when exploring such scenarios.
Regarding manufacturing emissions, the same extrapolation yields rough estimate of about 220 and 115 ktCO2e/year (for a 15 years lifespan). Contrary to energy consumption, these numbers are much larger than our estimates of the manufacturing emissions of the shared network equipment.

\subsection{Simplifications and design choices}
Our network and robot deployment models make several simplifications. We organized their discussion into three categories: possibly \textit{optimistic}, possibly \textit{pessimistic}, and \textit{unknown}.

\subsubsection*{Pessimistic choices}

On the rather pessimistic side, so far we considered the infrastructure of a single operator. Considering the mostly overlapping cells of the other three operators is expected to noticeably improve the percentage of manageable UAA for both the initial and upgraded configurations. However, it remains unclear whether the corresponding incremental energy and carbon footprints would increase proportionally. Accounting for multiple operators also raises the critical question of how to allocate robots across their different networks.
Such an extension is left as future work.

%
Just like~\cite{TheShiftProject_EnergieClimatReseaux_2024}, our power consumption model ignores sleep modes. Accounting for these modes during periods of no field-labour would decrease the static consumption part, especially for the cells that are explicitly added to serve robots.
%
Our scenarios make the conservative worst-case assumption that, within any sector, there is a moment of time where all required robots are simultaneously active, each with the same data bitrate need. In a more realistic context, the \textit{Stream} scenarios might not continuously require high-quality video streaming for all robots at all times. More generally, one could also expect that many sectors actually comprise heterogeneous labor agendas, thereby mitigating the peak of robot activity.

Last but not least, the average propagation radius and spectral efficiency values used in this study are representative of common smartphone devices. Perhaps it would be possible to increase both through the use of larger and more powerful antennas.

\subsubsection*{Optimistic choices}

The aforementioned pessimistic simplifications are counterbalanced by several optimistic ones.

For instance, our capacity model does not make use of any margin whereas it is classical to consider a 50\% margin to accommodate for the bursty nature of traffic, transmission error corrections, and other hazards~\cite{TheShiftProject_EnergieClimatReseaux_2024}.

Our model also assumes an homogeneous average capacity across each cell, whereas in reality, the maximum peak-rate drastically decreases with the distance between the cell's antenna and the mobile device. This means the maximum propagation radius should be reduced as the per-robot bitrate demand increases~\cite{Coupechoux_Performances5GEtude_2021}. Another consequence is that the actual available capacity depends on the relative location of the robots: a robot at the border of the cell might require the whole resources to maintain a high bitrate, hence reducing even more the practical propagation radius.

Moreover, maintaining low latency transfer usually implies dedicating more resource blocks to anticipate and reduce transmission hazards and errors, which in turn reduces the actual bitrate capacity.
\subsubsection*{Unknown choices}
A few other simplifications have effects that cannot be easily anticipated.

For instance, using guessed centers to determine the spatial extent of the cells is expected to lead to doubtful pixel-cell pairing where the inter-site distances is lower than half the propagation radius. Whereas such miss-pairing would yield to erroneous results at the scale of a single sector, we expect them to balance out at the global scale.

Similarly, our model assumes uniform propagation radius all over the territory. In practice, signal propagation is affected by many factors such as the local relief, forests, buildings, height of the antenna, etc. Again, such a simplification is important at a local scale, but we hope they compensate each other at the global scale.

Both shortcomings could be addressed through a more advanced propagation model or through empirical measurements.

We also made the assumption of a uniform bandwidth percentage (20\%) for other usages. In reality this load is expected to strongly vary over the territory, and it is unclear whether adjusting this percentage locally for each cell (either from empirical data or by exploiting population density and main roads) would significantly affect our global results.

At last, our network impact assessment model works in a static manner, without a progressive and timely simulation. This means that our modeled infrastructure is representative of the efficiency performance of today mobile network equipment ignoring the presence of older less efficient one, but also ignoring the possible future efficiency improvements that could be beneficial to the renewed and additional equipment involved in our scenarios. Nevertheless, this simplification is expected to be negligible when performing relative comparisons as in \sectref{Analysis}.

\subsection{Scenarios and future work}
On a broader level, we experimented our model on a single territory. It would be interesting to evaluate other territories with a perhaps different mobile network maturity, different agricultural parcels distributions, and another electricity mix.

Our scenarios simulate a homogeneous deployment of a single type of robot per scenario. A more realistic implementation could allow heterogeneous types of robots deployed together over UAA depending on characteristics such as the amount of UAA to be processed locally, or the available existing bandwidth.

Our scenarios do not incorporate potential benefits of applying robots to crops. This limitation should be addressed in future works by extending the scope of our study to balance network usages with a costs/benefits analysis of the effects of the application of different robot designs.

This study only considered existing network sites. A natural extension would be to design a site addition mechanism enabling to reach a given target of manageable UAA up to 100\%. Such an extension would provide more complete and representative insights regarding the impacts of these scenarios on the network. Note that a naive extrapolation based on our current incremental results and respective manageable UAA percentages would be erroneous because 1) such additional sites would exhibit a significantly higher incremental manufacturing and static power consumption footprints because all site equipment would have to be accounted for, and 2) the density and fragmentation level of the manageable UAA might not be representative of the remaining non manageable one.

\section{Conclusion}
We presented a methodology to assess the potential consequences of a large-scale deployment of agricultural robots regarding mobile network uses and scale. Different robotic scenarios were analyzed, based on existing and prospective applications. We compared possible deployments for both an existing network and an upgraded version. We discussed the static and dynamic energy consumption as well as the carbon footprint related to the use of mobile networks by robots. We also put them in perspective with regards to the agricultural area managed by simulated deployments.
Our findings indicate that even without adding new sites, the most data-intensive scenarios already exhibit significant additional impacts while managing only a small part of agricultural parcels.
Expanding the amount of manageable agricultural parcels through additional sites is expected to exacerbate these impacts.
These observations raise critical questions regarding both the technical feasibility and the relevance of deploying data-intensive digital agriculture applications at a large scale, in a context of environmental crisis where the ICT sector, like other industries, is expected to reduce its carbon emissions~\cite{FreitagEtAl_RealClimateTransformative_2021}.
For moderately data-intensive scenarios, further studies are required to get more thorough and representative insights. Such future work directions includes studying a site addition mechanism to reach higher levels of actual coverage, taking into account multiple operators, integrating empirical measurements, or also assessing a larger perimeter including servers and the robots themselves.

\bibliographystyle{apalike}
\bibliography{bibliography}

\begin{thebibliography}{}

\bibitem[{ADEME}, 2023a]{ADEME_negaoctet_2023}
{ADEME} (2023a).
\newblock Base-impact - digital services.
\newblock https://base-empreinte.ademe.fr/documentation/base-impact.

\bibitem[{ADEME}, 2023b]{ADEME_Electricite2023Mix_2023}
{ADEME} (2023b).
\newblock Electricit{\'e}/2023 - mix moyen/consommation.
\newblock https://base-empreinte.ademe.fr/donnees/jeu-donnees.

\bibitem[{Agence Nationale des Fréquences},
  2023]{AgenceNationaledesFrequences_DonneesInstallationsRadioelectriques_2023}
{Agence Nationale des Fréquences} (2023).
\newblock {Données sur les installations radioélectriques de plus de 5
  watts}.
\newblock
  https://www.data.gouv.fr/en/datasets/donnees-sur-les-installations-radioelectriques-de-plus-de-5-watts-1/.

\bibitem[{Agreste}, 2024]{Agreste_EnquetePratiquesCulturales_2024}
{Agreste} (2024).
\newblock Enquête pratiques culturales en grandes cultures 2021.
\newblock
  https://www.agreste.agriculture.gouv.fr/agreste-web/disaron/Chd2413/detail/.

\bibitem[AgroTIC, 2021]{AgroTIC_5GAgriculture_2021}
AgroTIC (2021).
\newblock {{5G}} et {{Agriculture}}.
\newblock Technical report, Chaire AgroTIC.

\bibitem[ARCEP, 2024a]{ARCEP_enquete_annuelle_2024}
ARCEP (2024a).
\newblock {Enquête annuelle pour un numérique soutenable}.

\bibitem[ARCEP, 2024b]{ARCEP_observatoire_2024}
ARCEP (2024b).
\newblock {Les services de communications électroniques en France 4e trimestre
  2023}.
\newblock Technical report, ARCEP.

\bibitem[Campbell et~al., 2017]{CampbellEtAl_AgricultureProductionMajor_2017}
Campbell, B.~M., Beare, D.~J., Bennett, E.~M., Hall-Spencer, J.~M., Ingram, J.
  S.~I., Jaramillo, F., Ortiz, R., Ramankutty, N., Sayer, J.~A., and Shindell,
  D. (2017).
\newblock Agriculture production as a major driver of the {{Earth}} system
  exceeding planetary boundaries.
\newblock {\em Ecology and Society}, 22(4):art8.

\bibitem[Coupechoux, 2021]{Coupechoux_Performances5GEtude_2021}
Coupechoux, M. (2021).
\newblock {Performances 5G: {\'e}tude compar{\'e}e en zones rurales et
  urbaines}.
\newblock Technical report.

\bibitem[Farmdroid, 2023]{Farmdroid_2023}
Farmdroid (2023).
\newblock {Farmdroid FD20}.
\newblock https://farmdroid.com/products/farmdroid-fd20/.

\bibitem[Freitag et~al., 2021]{FreitagEtAl_RealClimateTransformative_2021}
Freitag, C., {Berners-Lee}, M., Widdicks, K., Knowles, B., Blair, G.~S., and
  Friday, A. (2021).
\newblock The real climate and transformative impact of {{ICT}}: {{A}} critique
  of estimates, trends, and regulations.
\newblock {\em Patterns}, 2(9):100340.

\bibitem[Gerhards et~al., 2024]{GerhardsEtAl_ComparisonSevenInnovative_2024}
Gerhards, R., Risser, P., Spaeth, M., Saile, M., and Peteinatos, G. (2024).
\newblock A comparison of seven innovative robotic weeding systems and
  reference herbicide strategies in sugar beet
  ({{{\emph{Beta}}}}{\emph{vulgaris subsp. vulgaris}} {{L}}.) and rapeseed
  ({{{\emph{Brassica}}}}{\emph{ napus}} {{L}}.).
\newblock {\em Weed Research}, 64(1):42--53.

\bibitem[Golard et~al., 2023]{GolardEtAl_EvaluationProjection4G_2023}
Golard, L., Louveaux, J., and Bol, D. (2023).
\newblock Evaluation and projection of {{4G}} and {{5G RAN}} energy footprints:
  The case of {{Belgium}} for 2020--2025.
\newblock {\em Annals of Telecommunications}, 78(5-6):313--327.

\bibitem[Green et~al., 2021]{GreenEtAl_MeasurementLatencyRealtime_2021}
Green, M., Mann, D.~D., and Hossain, E. (2021).
\newblock Measurement of latency during real-time wireless video transmission
  for remote supervision of autonomous agricultural machines.
\newblock {\em Computers and Electronics in Agriculture}, 190:106475.

\bibitem[{GSMA}, 2020]{GSMA_SmartFarmingWeed_2020}
{GSMA} (2020).
\newblock Smart {{Farming}}: {{Weed Elimination}} with {{5G Autonomous
  Robots}}.
\newblock Technical report.

\bibitem[{Institut National De L'Information G{\'e}ographique Et
  Foresti{\`e}re},
  2022]{InstitutNationalDeLInformationGeographiqueEtForestiere_RegistreParcellaireGraphique_2022}
{Institut National De L'Information G{\'e}ographique Et Foresti{\`e}re} (2022).
\newblock Registre {{Parcellaire Graphique}}.

\bibitem[La~Rocca et~al., 2024]{LaRoccaEtAl_EstimatingCarbonFootprint_2024}
La~Rocca, P., Guennebaud, G., Bugeau, A., and Ligozat, A.-L. (2024).
\newblock Estimating the carbon footprint of digital agriculture deployment:
  {{A}} parametric bottom-up modeling approach.
\newblock {\em Journal of Industrial Ecology}, page jiec.13568.

\bibitem[Marwala, 2024]{Marwala2024}
Marwala, T. (2024).
\newblock {\em Embedded Versus Edge Versus Cloud Computing}, pages 171--187.
\newblock Springer Nature Singapore.

\bibitem[Moysiadis et~al., 2021]{MoysiadisEtAl_SmartFarmingEurope_2021}
Moysiadis, V., Sarigiannidis, P., Vitsas, V., and Khelifi, A. (2021).
\newblock Smart {{Farming}} in {{Europe}}.
\newblock {\em Computer Science Review}, 39:100345.

\bibitem[{Nperf}, 2023]{Nperf_BarometreConnexionsInternet_2023}
{Nperf} (2023).
\newblock Baromètre des connexions internet mobiles en france métropolitaine
  - année 2022.
\newblock
  https://media.nperf.com/files/publications/FR/2023-01-10\_Barometre-connexions-mobiles-metropole-nPerf-2022.pdf.

\bibitem[Oliveira et~al., 2021]{OliveiraEtAl_AdvancesAgricultureRobotics_2021}
Oliveira, L. F.~P., Moreira, A.~P., and Silva, M.~F. (2021).
\newblock Advances in {{Agriculture Robotics}}: {{A State-of-the-Art Review}}
  and {{Challenges Ahead}}.
\newblock {\em Robotics}, 10(2):52.

\bibitem[{Orphéon}, 2018]{Orpheon_CentreCalculOrpheon_2018}
{Orphéon} (2018).
\newblock {Le centre de calcul Orphéon}.
\newblock
  https://reseau-orpheon.fr/le-reseau-orpheon/infrastructure-centre-calcul/.

\bibitem[{Orphéon}, 2024]{Orpheon_RobotsRTKGuidage_2024}
{Orphéon} (2024).
\newblock {Robots RTK : guidage pr{\'e}cis et autonomie}.
\newblock https://reseau-orpheon.fr/robots-rtk-guidage-precis-et-autonomie/.

\bibitem[{Pixelfarming Robotics}, 2022]{PixelfarmingRobotics_2022}
{Pixelfarming Robotics} (2022).
\newblock {Robot One}.
\newblock https://pixelfarmingrobotics.com/robot-one/.

\bibitem[Ray et~al., 2019]{RayEtAl_ClimateChangeHas_2019a}
Ray, D.~K., West, P.~C., Clark, M., Gerber, J.~S., Prishchepov, A.~V., and
  Chatterjee, S. (2019).
\newblock Climate change has likely already affected global food production.
\newblock {\em PLOS ONE}, 14(5):e0217148.

\bibitem[Roberts and Pecka, 2018]{RobertsPecka_4GNetworkPerformance_2018}
Roberts, A. and Pecka, A. (2018).
\newblock {{4G}} network performance analysis for real-time telemetry data
  transmitting to mobile agricultural robot.
\newblock In {\em 17th {{International Scientific Conference Engineering}} for
  {{Rural Development}}}.

\bibitem[Rouphael et~al., 2023]{RouphaelEtAl_ImpactNetworksGreenhouse_2023}
Rouphael, R.~B., Selva, E., Huang, Y., Mouawad, N., Boudry, G., Parmantier, X.,
  and Gati, A. (2023).
\newblock The {{Impact}} of {{Networks}} in the {{Greenhouse Gas Emissions}} of
  a {{Major European CSP}}.
\newblock In {\em {{International Conference}} on {{Electrical}}, {{Computer}}
  and {{Energy Technologies}}}.

\bibitem[Ruigrok et~al.,
  2020]{RuigrokEtAl_ApplicationSpecificEvaluationWeedDetection_2020}
Ruigrok, T., Van~Henten, E., Booij, J., Van~Boheemen, K., and Kootstra, G.
  (2020).
\newblock Application-{{Specific Evaluation}} of a {{Weed-Detection Algorithm}}
  for {{Plant-Specific Spraying}}.
\newblock {\em Sensors}, 20(24):7262.

\bibitem[Sparrow and Howard,
  2021]{SparrowHoward_RobotsAgricultureProspects_2021}
Sparrow, R. and Howard, M. (2021).
\newblock Robots in agriculture: Prospects, impacts, ethics, and policy.
\newblock {\em Precision Agriculture}, 22(3):818--833.

\bibitem[{The Shift Project}, 2024]{TheShiftProject_EnergieClimatReseaux_2024}
{The Shift Project} (2024).
\newblock {Energie, climat : des r{\'e}seaux sobres pour des usages
  connect{\'e}s r{\'e}silients}.
\newblock Technical report, The Shift Project.

\bibitem[Tomaszewski et~al.,
  2022]{TomaszewskiEtAl_ApplicationMobileNetworks_2022}
Tomaszewski, L., Ko{\l}akowski, R., and Zag{\'o}rda, M. (2022).
\newblock Application of {{Mobile Networks}} ({{5G}} and {{Beyond}}) in
  {{Precision Agriculture}}.
\newblock In {\em Artificial {{Intelligence Applications}} and
  {{Innovations}}}.

\bibitem[Zhivkov et~al., 2023]{ZhivkovEtAl_5GFarmEvaluating_2023}
Zhivkov, T., Sklar, E.~I., Botting, D., and Pearson, S. (2023).
\newblock {{5G}} on the {{Farm}}: {{Evaluating Wireless Network Capabilities}}
  and {{Needs}} for {{Agricultural Robotics}}.
\newblock {\em Machines}, 11(12):1064.

\bibitem[Zoom, 2024]{_ZoomSystemRequirements_}
Zoom (2024).
\newblock {Zoom system requirements: iOS, iPadOS, and Android}.

\end{thebibliography}

\end{document}